\documentclass[iop]{emulateapj}
\usepackage{graphicx}
\usepackage{amssymb}
\usepackage{amsmath}
\usepackage{natbib}
\usepackage[breaklinks,colorlinks=true,urlcolor=blue,citecolor=blue,linkcolor=blue]{hyperref}

\shorttitle{Laboratory measurements of the 'dark matter' $\sim$3.5 $\mathrm{ke}$V X-ray line}
\shortauthors{Shah et al.}

\begin{document}

	\title{Laboratory measurements compellingly support charge-exchange mechanism for the 'dark matter' $\sim$3.5 \MakeLowercase{ke}V X-ray line}
	
	\author{Chintan Shah\altaffilmark{1,\,$\ast$}, Stepan Dobrodey\altaffilmark{1}, Sven Bernitt\altaffilmark{1,\,2}, Ren\'{e} Steinbr\"{u}gge\altaffilmark{1,\,$\dagger$}, Jos\'{e} R. Crespo L\'{o}pez-Urrutia\altaffilmark{1}, \\
	Liyi Gu\altaffilmark{3}, Jelle Kaastra\altaffilmark{3,\,4}}
	\affil{$^1$Max-Planck-Institut f\"ur Kernphysik, Heidelberg, 69117 Heidelberg, Germany}
	\affil{$^2$Institut f\"ur Optik und Quantenelektronik, Friedrich-Schiller-Universit\"at, 07743 Jena, Germany}
	\affil{$^3$SRON Netherlands Institute for Space Research, Sorbonnelaan 2, 3584 CA, Utrecht, The Netherlands}
	\affil{$^4$Leiden Observatory, Leiden University, PO Box 9513, 2300 RA Leiden, The Netherlands}
	\altaffiltext{$\ast$}{chintan@mpi-hd.mpg.de}
	\altaffiltext{$\dagger$}{Now at Cyclotron Institute, Texas A\&M University, College Station, TX 77843, USA} 	


\begin{abstract}
	The reported observations of an unidentified X-ray line feature at $\sim$3.5 keV have driven a lively discussion about its possible dark matter origin. Motivated by this, we have measured the \emph{K}-shell X-ray spectra of highly ionized bare sulfur ions following charge exchange with gaseous molecules in an electron beam ion trap, as a source of or a contributor to this X-ray line. We produce $\mathrm{S}^{16+}$ and $\mathrm{S}^{15+}$ ions and let them capture electrons in collision with those molecules with the electron beam turned off while recording X-ray spectra. We observed a charge-exchanged-induced X-ray feature at the Lyman series limit (3.47 $\pm$ 0.06 keV). The inferred X-ray energy is in full agreement with the reported astrophysical observations and supports the novel scenario proposed by Gu and Kaastra (A \& A \textbf{584}, {L11} (2015)).\\
\end{abstract}


%
%
\section{Introduction}
%
%

Recently reported observations of a weak X-ray emission line-like feature at an energy of approximately $\sim$3.5 keV from the Perseus galaxy cluster, the M31 galaxy~\citep{boyarsky2014} and in stacked spectra of 73 galaxy clusters~\citep{bulbul2014} with the X-ray cameras of the \textit{XMM-Newton} telescope have attracted enormous attention. The reason for this is the lack of an immediate identification of the feature in standard wavelength tables. Although the existence of a very large number of unidentified transitions in all spectral ranges is a well-known fact, it has been widely assumed that our knowledge of the X-ray emission spectra from atomic sources was sufficiently well modeled to pinpoint this transition as an exceptional phenomenon. In particular, speculations about a possible dark matter origin of this observed X-ray line feature from galaxy clusters have sparked an incredible interest in the scientific community and given rise to a tide of publications attempting to explain the possible reason for the observed unidentified line feature (ULF). Among other possibilities, the origin of this line has been hypothesized as the result of decaying, long-sought dark matter particle candidates -- sterile neutrinos, presumably based on the fact that this X-ray line is not available in the present atomic databases for thermal plasmas~\citep{bulbul2014,foster2012}. 

Similar signals were later detected from the Galactic Center~\citep{boyarsky2015} and from the Perseus cluster core with the help of the \textit{Suzaku}~\citep{urban2015} telescope. While these studies were able to establish upper flux limits for the ULF, they could not provide conclusive evidence for it due to statistical and model uncertainties. A very recent study on the dwarf spheroidal galaxy Draco from observations with~\textit{XMM-Newton} rules out a possible dark matter decay origin of the ULF~\citep{jeltema2016} based on an incompatibility with the expected dark matter distribution of that system. A comprehensive search of ULF in stacked galaxy spectra by~\citet{anderson2015} concluded similar with no significant evidence of any emission line at 3.5 keV.

Given the importance of the matter, a very careful spectral analysis should be carried out in order to first exclude all possible known causes for X-ray emissions in this spectral range. Unfortunately, the standard spectral databases used for comparison and the models based upon them have in part to rely on atomic structure calculations, since the body of laboratory data on X-ray emission lines is by far not complete~\citep{beiersdorfer2003a}. In this work, we show experimental data which strongly support the cautious explanation of the ULF recently given by~\citet{gu2015}: this intriguing X-ray line feature arises from charge exchange between fully stripped sulfur ions and atomic hydrogen, populating states in high principal quantum numbers of the subsequently formed hydrogenlike sulfur ions. In this model, it is compelling that X-rays should be emitted at $\sim$3.5 keV by a set of $\mathrm{S}^{15+}$ transition from $n\geq9$ to the ground state, where $n$ is the principal quantum number. This scenario has to be considered since the highly ionized plasma present in galaxy clusters certainly contains $\mathrm{S}^{16+}$ and $\mathrm{S}^{15+}$ ions~\citep{mushotzky1981}. 

%
%
\section{Charge exchange excitation of X-ray transitions}
%
%

Charge exchange (hereafter CX) occurs when a neutral atom collides with a sufficiently charged ion, which becomes typically an excited product species. For a highly charged ion (HCI) moving at thermal velocities, the electron transfer takes place at --atomically speaking-- large distances, since the potential well of the neutral donor is strongly distorted by the approaching HCI at internuclear separations of several atomic units. The lowering of the potential barrier between the two charged centers leads to the formation of a short-lived quasi-molecular state for the outermost electrons of the donor, with the electronic wavefunction extending between the two centers, the projectile HCI, and the donor. In a classical picture of a collision at thermal energies, the electron repeatedly visits both between the approaching and receding phase of the collision and is likely captured by the center with the higher charge after the passage. This process is usually be modeled using e.~g.~the so-called over-the-barrier method (OBM) of~\citet{ryufuku1980}, and more advanced models liked the extended OBM of~\citet{niehaus1986}, multi-channel Landau-Zener (MCLZ)~\citep{cumbee2016} methods and various other ones. Charge exchange between HCI and neutral donors leads to the formation of a high Rydberg state in the down-charged HCI projectile for the range of thermal collision energies. The cross sections for populating such states are rather large ({$10^{-15}\,\mathrm{cm}^{2}$} and higher for HCI with charges 10-20). Thereafter, radiative decay of those excited states with high principal quantum number $n=n_{\mathrm{cx}}$ to the ground state $n=1$ produces X-ray line emission. If the colliding HCI has \emph{K}-shell vacancies, it will fill each of them through emission of an X-ray photon. In the case of multiple electron capture from a many-electron neutral donor, various non-radiative Auger processes can also lead to relaxation of the system~\citep{fischer2002,knoop2008,xue2014}.

Following the discovery of X-rays from comets~\citep{lisse1996,lisse2001,dennerl1997} and its explanation as being caused by the interaction of solar wind HCI and neutral coma gas~\citep{cravens1997}, very clear observations of CX-induced X-ray emission have provided insights into the interaction between cold and hot astrophysical plasmas. The pioneering reproduction in the laboratory of the low resolution spectra observed in comets by~\citet{beiersdorfer2003} showed how CX-induced X-ray emission can be used to diagnose the properties of the solar wind dynamics. Absolute CX cross sections were also obtained in the laboratory for solar-wind HCI interacting with the neutrals present in comets~\citep{greenwood2000a,greenwood2000,mawhorter2007}. Due to the HCI abundance, CX is a ubiquitous process which should always be considered wherever hot plasma interacts with a neutral medium. This is obviously the case in galaxy clusters where hot intracluster medium interacts with cold clouds dwelling around the central galaxies~\citep{gu2015}. 

Laboratory studies have been carried out by several groups with various methods (see e.~g.~\citep{dijkkamp1985,janev1983,hoekstra1990,bodowits2004,bodowits2006,trassinelli2012}). Moreover, CX has been recognized as essential for understanding the ionization equilibrium of laboratory plasmas, and it also affects the storage time in ion traps and the energy transport mechanism in the edge region of tokamak fusion plasmas~\citep{beiersdorfer2000,leutenegger2010}. This has motivated a number of CX studies at the LLNL electron beam ion trap (EBIT) group~\citep{beiersdorfer2000,beiersdorfer2003,wargelin2005,leutenegger2010,martinez2014} and other laboratories~\citep{allen2008,fischer2002,knoop2008,xue2014}, with X-ray spectral analysis carried out at both low and high resolution. Various theoretical models have been invoked for data analysis. While confirming the role of CX, those experiments have nonetheless pointed out discrepancies in the quantitative modeling of the relative intensity of CX-fed transitions that are still a matter of active research in both theory and experiment.

\begin{figure}
\includegraphics[width=0.95\columnwidth]{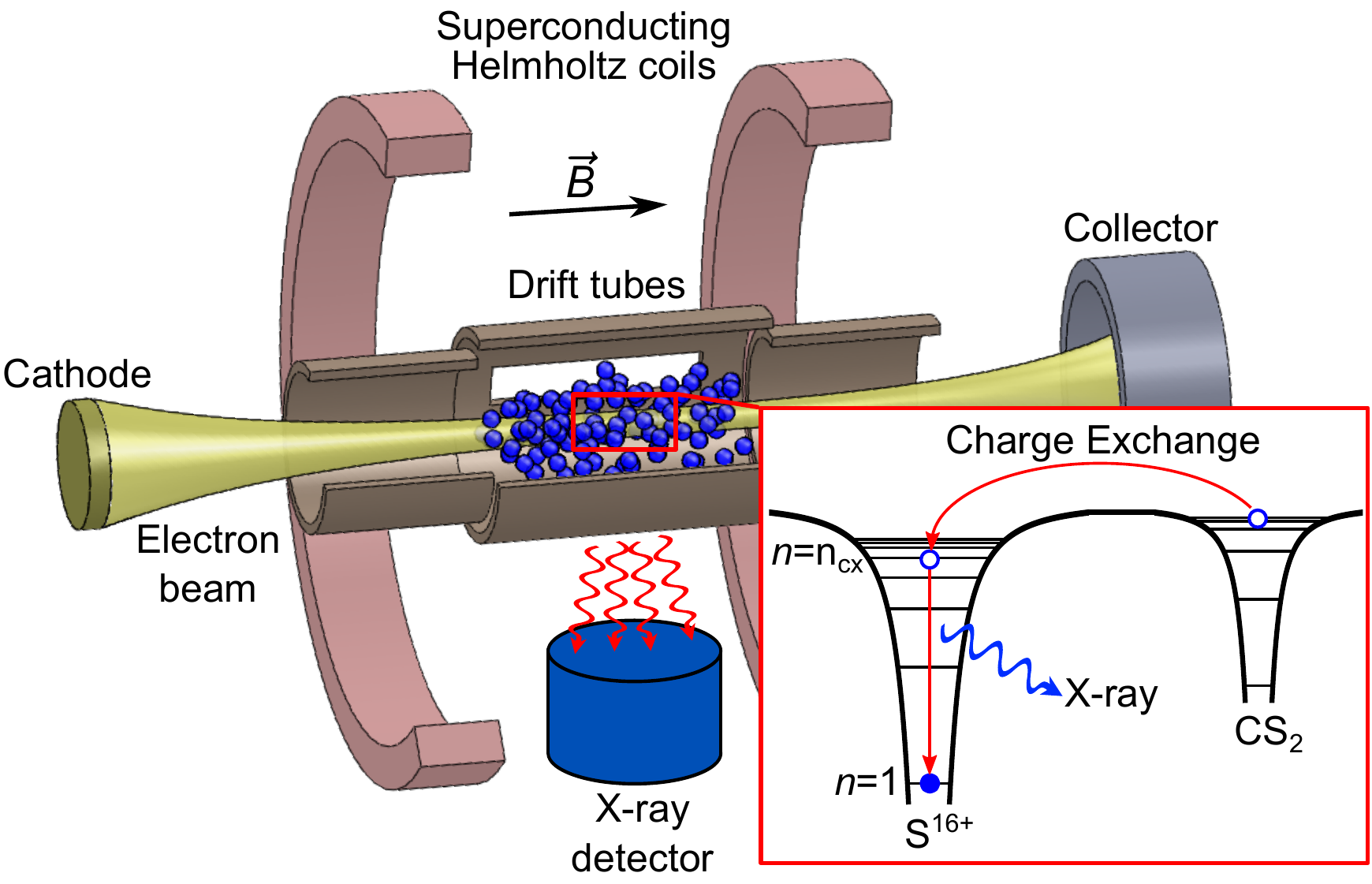}
\caption{Schematic drawing of an electron beam ion trap. The electrons emitted from the cathode are accelerated towards the trap center where they collide with CS$_2$ molecules. Highly charged sulfur ions are produced and trapped within the electron beam. During the beam-off mode (inset figure), the magnetically trapped sulfur ions interact with neutrals through charge exchange and emit X-rays.~\label{fig:ebit}}
\end{figure}

A common feature of most X-ray observations is the lack of sufficient resolution capable of distinguishing the principal quantum number $n$ and angular momentum quantum numbers $l$ of the state in which the electron is captured. This leads to the use of various definitions of "hardness" for the observed spectra, which are basically intensity ratios ($\mathcal{R}$ and $\mathcal{H}$ respectively for hydrogenlike and heliumlike lines) between the unresolved transitions from $n\geq 3\rightarrow 1$ to the better separated transition $n=2\rightarrow 1$. A few high resolution experiments using X-ray microcalorimeters~\citep{leutenegger2010} or crystal spectrometers~\citep{rosmej2006} do not suffer from these limitations which strongly hinder astrophysical studies. Strong expectations in the performance of \emph{Astro-H}, which among other things would have provided novel means to understand the contributions of this process have been frustrated by the untimely demise of that critical mission.

Among other findings, the studies have revealed some intricacies in the dependence of the range of $nl$ distribution on the donor neutral which are still not fully understood~\citep{beiersdorfer2000,beiersdorfer2001} and can lead to an uncertainty in the modeling of the spectral hardness. One particularly striking example~\cite{leutenegger2010} showed how simultaneously prepared Ar$^{17+}$ and P$^{15+}$ ions in spite of their similarity of their electronic structure, produced X-ray spectra that allowed to infer rather disparate $nl$ populations in the capture process. Further, studies have shown effects of simultaneous and sequential multiple electron capture processes in collisions~\citep{otranto2011,otranto2014}. These are particularly relevant for HCI interactions with a comet, where many-electron donor species are present~\citep{ali2005,ali2010,ali2016}, while in interaction with atomic hydrogen single capture is the only possible process.

Motivated by this, we tested the hypothesis of the CX scenario proposed by~\citet{gu2015} by investigating in the laboratory the X-ray emission spectra of fully stripped sulfur ions after interaction with neutral gas. In the following sections, we will discuss the experimental setup, data analysis and results that corroborate the theoretical predictions of Gu and astrophysical observations.

%
%
\section{Experimental technique}
%
%

For the measurements, we use the magnetic trapping mode of an EBIT~\citep{beiersdorfer2000}. The experimental principle is the following: Bare $\mathrm{S}^{16+}$ and H-like $\mathrm{S}^{15+}$ ions are produced in an EBIT by interaction with an intense electron beam; recombination with a neutral species gas is observed after turning off the electron beam while keeping the HCI magnetically confined~\citep{beiersdorfer2000}. By differentially varying the production conditions of the two ions, we distinguished their respective CX contributions to the recorded X-ray spectra with a photon energy resolution close to that of the X-ray cameras on board of \textit{XMM-Newton}. Under a broad range of conditions, a 3.5 keV transition clearly shows up in the spectra which could well explain the aforementioned astrophysical observations. 

We carried out the experiment with FLASH-EBIT~\citep{epp2007,epp2010}, a device used primarily for soft and hard X-ray excitation of HCI~\citep{bernitt2012} with free-electron lasers and synchrotron radiation at its base location in the Max-Planck-Institute for Nuclear Physics, Heidelberg. The setup is sketched in Fig.~\ref{fig:ebit}. An intense, magnetically-compressed electron beam ionizes neutrals in its path and subsequently traps the ions in tight orbits around its own propagation axis by means of its negative space charge potential. Successive electron impact ionization occurs by the monoenergetic electron beam, which is focused to approximately 25\,$\mu$m radius by a coaxial 6\,T fields generated by a superconducting Helmholtz coil setup. The ions are trapped along the beam axis inside a potential well ($\sim q\times100\,\mathrm{V}$, where $q$ is the ionic charge) produced by a set of drift tubes. Radial confinement is essentially due to the negative space charge potential of the electron beam (with a depth similar in magnitude to the axial one), but also by the axial magnetic field. In this way, an ion cloud containing millions of HCI with an extension of 50\,mm length and less than 1\,mm in diameter forms in a fraction of a second. It is crucial for this experiment that even after turning off the electron beam, the magnetic field efficiently confines the beam-generated HCI within a cylindric volume of larger diameter than under electron beam trapping, the so-called magnetic trapping mode~\citep{beiersdorfer2000}.    

Sulfur atoms were continuously brought into the interaction and trapping region using a volatile molecular compound, carbon disulfide (CS$_2$), forming an extremely tenuous molecular beam intersecting the electron beam. This molecular beam conveniently serves both for the production of sulfur HCI and as a donor species for CX. For the production of sulfur HCI in the charge states of interest, the electron beam energy has to exceed the ionization potential of heliumlike $\mathrm{S}^{14+}$. The stability of this type of closed-shell ions makes it necessary to raise the collision energy well above that threshold value. While this is advantageous in terms of the yield of $\mathrm{S}^{15+}$ and $\mathrm{S}^{16+}$ ions, it also implies that the three charge states can be simultaneously present in the trap. In order to distinguish the contributions of the two species giving rise to transitions to $n=1$, we perform a differential measurement by changing the beam energy below and above the threshold for production of bare $\mathrm{S}^{16+}$ ions.   
\begin{figure}
	\includegraphics[width=0.95\columnwidth]{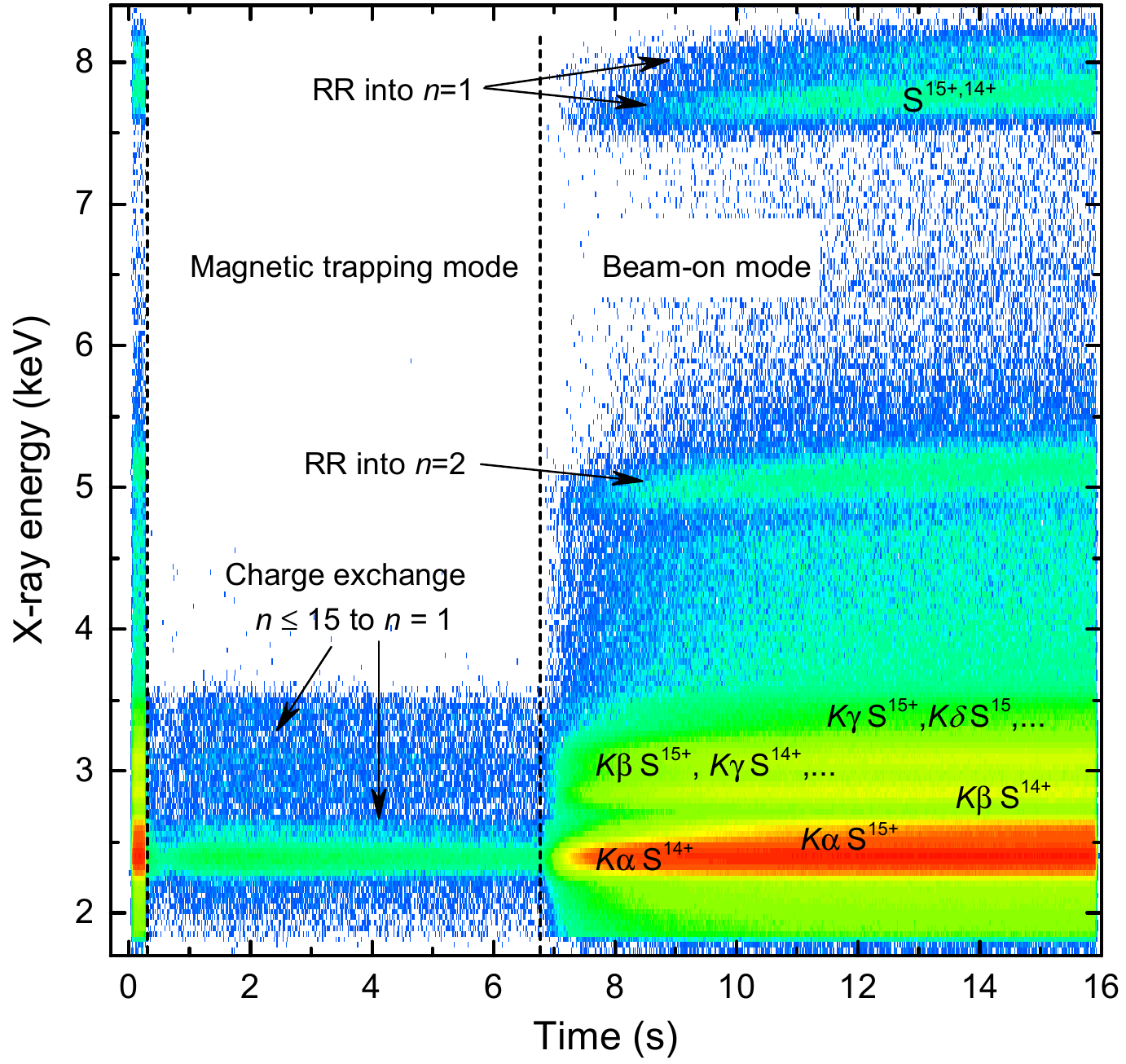}
	\caption{X-ray spectrum during beam-on and beam-off periods (magnetic trapping mode) using an electron beam energy of $4.6\,\mathrm{keV}$ and an injection pressure of $\sim 10^{-8}\,\mathrm{mbar}$, corresponding to an estimated neutral density of $\sim 3\times10^{-6}$ particles per cm$^{-3}$ at the interaction region, where they collide with deeply trapped HCI at an estimated translational temperature of $\sim$800\,eV ($\sim$9\,MK). The radiative recombination of the monoenergetic 4.6\,keV beam electrons into $n=1, 2$ states of $\mathrm{S}^{16+..14+}$ is visible at $\sim$8 and 5\,keV, respectively.\label{fig:cx_2d}}
\end{figure}

The technique requires cyclic, rapid switching of the electron beam: beam-on mode to produce and trap HCI, and beam-off mode confining the ions only by the magnetic field in the absence of electron-impact excitation. With this technique, X-ray emission from charge exchange can be distinguished from the copious X-ray productions while the beam is on. The electron beam carries a current of 150~mA during the production sub-cycle lasting for 9.4~s, after which the electron beam is turned off for 6.6~s, see Fig.~\ref{fig:cx_2d}. Switching is performed by controlling the electron gun with a fast high-voltage amplifier driven by a periodical rectangular signal, leading to an overall duty cycle of $\sim40\,\%$. A synchronous linear time ramp allows timing of the detected X-ray photons within the cycle. These are recorded with a commercial windowless silicon drift detector with an energy resolution of FWHM $\sim 150\,\mathrm{eV}$. While this detector does not offer the excellent photon energy resolution afforded by an X-ray microcalorimeter as used in CX experiments at LLNL~\citep{beiersdorfer2003,leutenegger2010}, it provides a conveniently large detection solid angle and reduces our measuring time.

%
%
\section{Results and data analysis}
%
%

A typical X-ray spectrum is shown in Fig.~\ref{fig:cx_2d}. During the beam-on mode, electron impact excitation into various~\textit{n}-states of $\mathrm{S}^{15+}$ and $\mathrm{S}^{14+}$ and subsequent decay to the ground state leads to the emission of \emph{K}-shell X-rays. Furthermore, monoenergetic electrons from the beam also radiatively recombine, predominantly into the $n=1$ and $n=2$ states, giving rise to the $\sim 8\,\mathrm{keV}$ and $\sim 5\,\mathrm{keV}$ transitions, respectively. Their photon energy is the sum of the electron beam energy and the ionization potential of the respective shell and ion.
In contrast, during the magnetic trapping mode the emission of X-rays is much weaker. After turning off the electron beam, metastable states populated by electron impact very quickly relax. In this regime, this happens for sulfur ions typically within microseconds, unresolvable in the present time scale. Some impurity HCI from barium emanating from the cathode are co-trapped and appear as weak features at 4.5~keV. The dominant emission during the beam-off period is due to the radiative relaxation of high Rydberg states that are populated by CX, resulting in X-ray transitions starting from principal quantum numbers $n_{\mathrm{cx}}=15,..,3 \rightarrow n=1$. Various radiative cascades also feed the $n=2$ state, producing a strong Lyman-$\alpha$ X-ray line at $\sim$2.4~keV.

\begin{figure}
	\includegraphics[width=0.9\columnwidth]{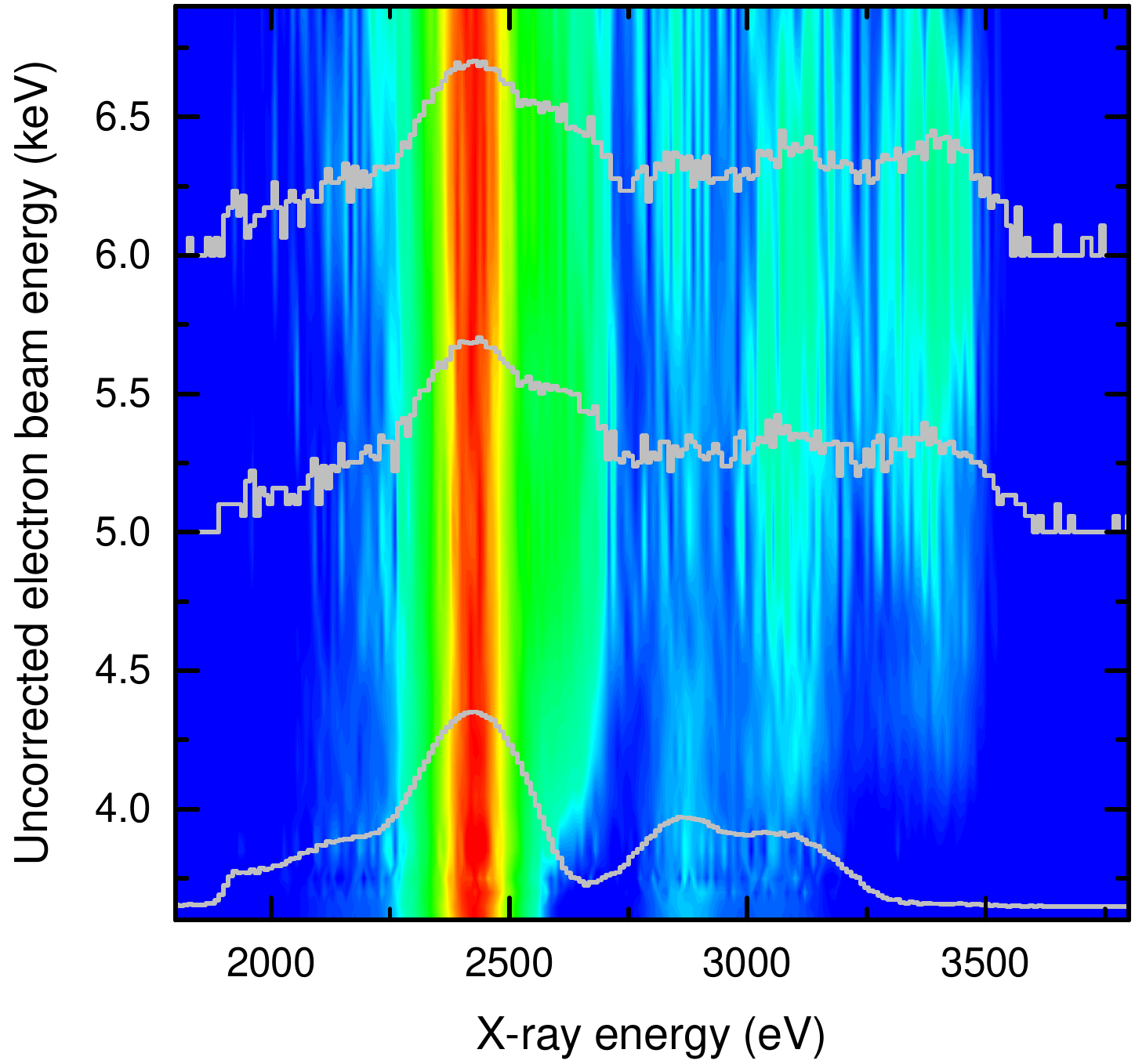}
	\caption{Charge exchange induced X-ray spectra (in magnetic trapping mode) as a function of electron beam energy. Electron beam energies are not corrected for the negative space charge of the electron beam that is approximately 450~eV at 4~keV of beam energy and 150~mA of beam current.\label{fig:scans}}
\end{figure}

We measured spectra at electron beam energies in the range from 3.4 to 7~keV and with $\mathrm{CS}_{2}$ injection pressures in the first stage of the differential pumping system used to form the molecular beam varying from $10^{-8}$ to $10^{-5}$~mbar. The scan of the electron beam energy shown in Fig.~\ref{fig:scans} was performed at an injection pressure of $\sim 10^{-8}\,\mathrm{mbar}$. Pressure scans were carried out with a fixed electron beam energy of $5\,\mathrm{keV}$. For bare sulfur production, a minimum electron beam energy of 3494~eV is required. Accordingly, the experiment shows a clear appearance and evolution of the $n_{\mathrm{cx}}\leq 15 \rightarrow n=1$ transitions at energies $3.30 - 3.48\,\mathrm{keV}$ at electron beam energy above 4000~eV. 

%

%
%
\subsection{X-ray energy calibration}
%
%


For the X-ray detector energy calibration, we recorded in subsequent measurements transitions excited by resonant dielectronic recombination (DR) into highly charged sulfur, argon and barium ions. Well known, dominant \emph{K}-shell X-ray transitions arising from DR into $\mathrm{S}^{14+}$ and $\mathrm{Ar}^{16+}$ ions, and the \emph{L}-shell lines due to direct excitation of $\mathrm{Ba}^{46+}$ from $n=2$ to $n=3$ were observed. Transition energies for those lines were obtained with the help of theoretical calculations performed with Flexible Atomic Code (FAC)~\citep{gu2008}. Previous experimental results of DR (e.~g., in~\citet{shah2015,shah2016}) were in good agreement with FAC predictions, making it sufficiently reliable for the present calibration. We estimate a theoretical uncertainty of approximately 10 eV due to the imperfectly known charge state distribution of the trapped ions. The analog-to-digital converter (ADC) signal histogram of each transition was fitted with a Gaussian to determine their ADC-channel positions precisely. A linear energy scaling factor of ~18.7 (eV per detector channel) with an offset of~116.7 channels with a relative calibration factor uncertainty of $\sim 0.3\%$ was determined. The offset was further corrected to the energy of the well-known Lyman-$\alpha$ line of hydrogenlike sulfur. Given the width of the observed features, the calibration is sufficiently accurate for the purpose of this work.

%
%
\subsection{Spectral fitting}
%

%
\begin{figure*}
	\includegraphics[width=0.97\textwidth]{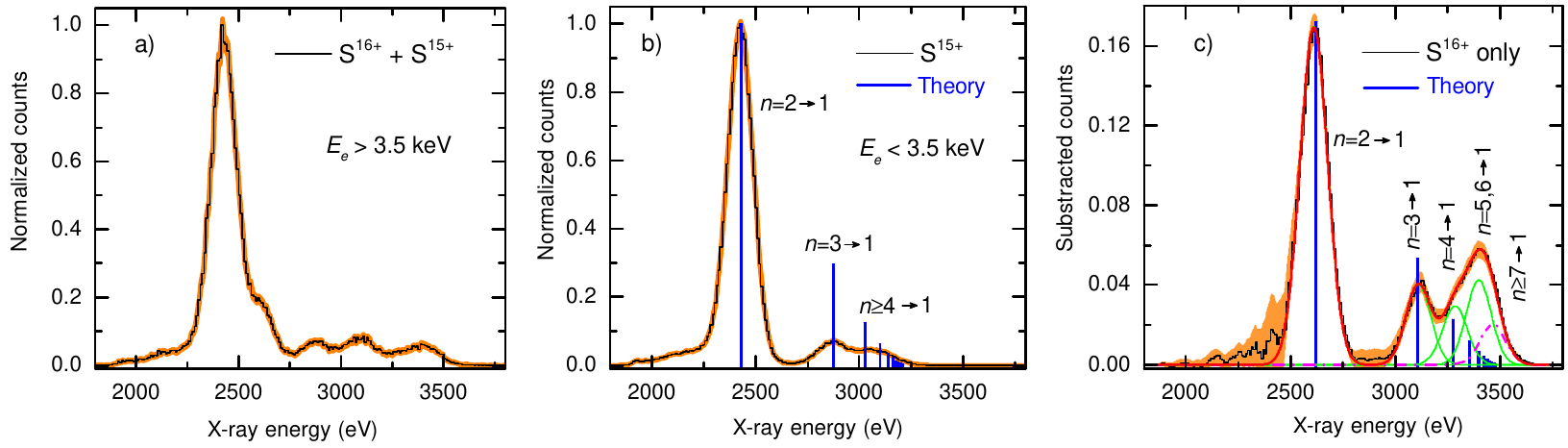}
	\caption{X-ray emission spectra for charge exchange of highly ionized sulfur with CS$_2$ injected into the trap. (a) an admixture of $\mathrm{S}^{16+}$ and $\mathrm{S}^{15+}$ ions at electron beam energy above the ionization threshold of bare sulfur, (b) $\mathrm{S}^{15+}$ below the ionization threshold of bare sulfur, and (c) subtracted spectrum with only $\mathrm{S}^{16+}$ contributions. The red curve represent the fit of the complete CX spectrum of bare sulfur, thin green curves correspond to individual fits to the CX-fed radiative transitions from different $n$ levels, and the dash-dotted curve represent $n\geq7\rightarrow1$ transition at 3.47 $\pm$ 0.06 keV. The theoretical predictions of their energies are shown by blue vertical lines.\label{fig:CX_spec}}
\end{figure*}

In Fig.~\ref{fig:CX_spec} we display the total number of detected photons within the magnetic trapping mode as a function of photon energy. To ensure that the spectra are completely free of photons emitted due to the electron beam-ion interaction, only those counts are considered that are detected 1~s (well above the beam ramp down time) after the electron beam was switched off in each measurement cycle. We extract results for $\mathrm{S}^{16+}$ and $\mathrm{S}^{15+}$ ions by varying the production conditions and thus changing the relative populations of the types of ions in order to distinguish their respective contributions. A pure spectrum of $\mathrm{S}^{15+}$ ions can be produced by choosing an electron beam energy below the ionization potential of this ion. For the reason mentioned above, above the ionization potential of bare sulfur, the emission spectrum is an admixture of $\mathrm{S}^{16+}$ and $\mathrm{S}^{15+}$ ions, see Fig.~\ref{fig:CX_spec} (a) and (b). We remove the contribution of $\mathrm{S}^{15+}$ ions to this mixed spectrum by subtracting the emission components due to the presence of $\mathrm{S}^{15+}$ ions, and obtain in this way the spectrum of $\mathrm{S}^{16+}$ ions, as shown in Fig.~\ref{fig:CX_spec} (c).

A radiative cascade network for the $\mathrm{S}^{16+}$ and $\mathrm{S}^{15+}$ ions was calculated with FAC. Transition energies and oscillator strengths were obtained for atomic levels up to $n=15$. The transition energies for $n_{\mathrm{cx}}=15,..,2 \rightarrow 1$ are presented in Fig.~\ref{fig:CX_spec} for both $\mathrm{S}^{16+}$ and $\mathrm{S}^{15+}$ ions. Energies were extracted by fitting Gaussian distributions to the experimental CX spectrum of bare sulfur. For the fitting, we treated centroids and amplitudes as free parameters and the widths were fixed to the known energy resolution of the detector (150~eV). Fits are represented by the red curve in the Fig.~\ref{fig:CX_spec} (c). In total five Gaussians were fitted to the spectrum and the typical reduced $\chi^2$ value was obtained to be 1.06 from the fits. The exact energy positions were extracted, respectively, 2.61 $\pm$ 0.001,~3.11 $\pm$ 0.005,~3.28 $\pm$ 0.02,~3.40 $\pm$ 0.044,~and~3.47 $\pm$ 0.06~keV correspond to $n=2\rightarrow1$, $n=3\rightarrow1$, $n=4\rightarrow1$, $n=5,\,6\rightarrow1$ and $n\geq 7\rightarrow1$. These values are in full agreement with the predicted line centroids with FAC, confirming the accuracy of our photon energy calibration. Several measured spectra were analyzed with the same technique; the CX-fed transition from $n_{\mathrm{cx}}\geq 7 \rightarrow n=1$ is observed in all the bare sulfur spectra at 3.47 $\pm$ 0.06 keV. Our results perfectly agree with the systematic microcalorimeter  measurements of \citet{martinez2014} at the LLNL EBIT, which used both He and SF$_6$ as donor gases. The high resolution spectra with He as CX donor show a very strong Lyman$_\eta$ at ~3438 eV, while with SF$_6$ as a donor the dominance of that transition is less pronounced. Also the hardness ratios $\mathcal{R}$ and $\mathcal{H}$ (for  the $\mathrm{S}^{15+}$ and $\mathrm{S}^{14+}$ transitions respectively) found in the present work for the CS$_2$ case agree very well with the SF$_6$ values of \citet{martinez2014}. Moreover, we also made a proof-of-principle CX study with~$\mathrm{Ar}^{18+}$ and $\mathrm{Ar}^{17+}$ ions. The results are consistent with the previous measurements by~\citet{beiersdorfer2000} and~\citet{allen2008}, and show similar discrepancy in relative intensity of CX induced transitions in comparison with the corresponding model.

%
%
\subsection{Comparison with CX modeling}
%
%

We now compare the CX spectra obtained for bare sulfur in the experiment with the calculations by~\citet{gu2015} as implemented in the plasma emission model in the SPEX package~\citep{kaastra1996}. As described in~\citet{gu2016}, the CX model incorporates reaction rates mainly calculated by the multi-channel Landau-Zener method assuming an atomic hydrogen target~\citep{cumbee2016}, and interpolates the rates when the actual atomic calculations are not available. Only single electron capture is considered in the model. Emission spectra are then modeled including radiative cascades based on atomic levels and transition-probability data which are complete up to $n=16$. Since our donor is not atomic hydrogen, but $\rm CS_{2}$ gas, we scale the value of $n$ for the most-populated level $n_{\rm p}$ according to Eq.3 of~\citet{gu2016}, $n_{\rm p, CS_{2}} = n_{\rm p, H} \times \sqrt{I_{\rm H}/I_{\rm CS_{2}}}$, where $n_{\rm p, H} = 10$ for the low energy limit, $I_{\rm H} =$ 13.6 eV and $I_{\rm CS_{2}} =$ 10.1 eV are the ionization potentials of atomic hydrogen and $\rm CS_{2}$, respectively. This yields for $n_{\rm p, CS_{2}} = 12$. 

\begin{figure}[b]
	\includegraphics[width=0.95\columnwidth]{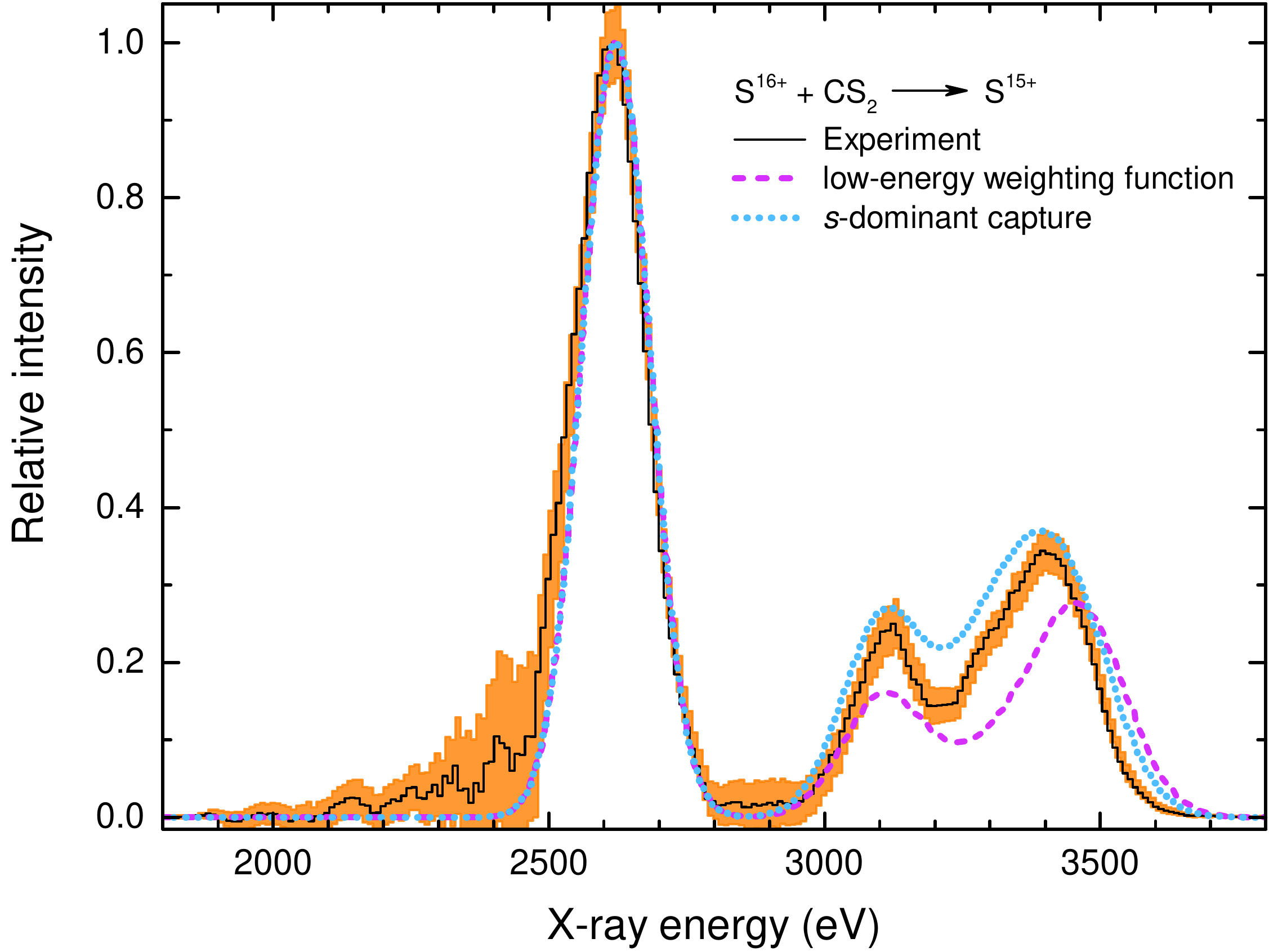}
	\caption{Synthetic CX spectra with typical low-energy weighting function (dashed line) and with relatively high capture into the $s$-state as reported in e.g.,~\citet{nolte2012} (dotted line) are compared with the experimental CX spectrum of bare sulfur with CS$_2$.	\label{fig:exp_th}}
\end{figure}

To model the sub-level splitting, we first consider a typical low-energy weighting function $W^{\rm l2}_{n} (l)$ (Eq.~5 of~\citet{gu2016}), which favors capture into low angular momentum $l = 1, 2$. This function is recommended for collision energies $< 10-100$ eV/u~\citep{krasnopolsky2004}. As shown in Fig.~\ref{fig:exp_th}, the model with $W^{\rm l2}_{n} (l)$ predicts a pileup of high-$n$ transitions, which slightly shifts from the data towards a higher energy value. We speculate that to get better agreement with the experimental peak, the capture cross section to $l=0$, which in turn cascades to $p-$levels at lower Rydberg states, must be enhanced. Then, following~\citet{mullen2016}, we construct an ad hoc $l-$distribution function $W^{\rm l2}_{n} (l^{\prime})$ in which $l^{\prime} = l - 1$. Such an $s$-dominant capture was reported in a few theoretical calculations, e.~g.,~\citet{nolte2012}. As shown in Fig.~\ref{fig:exp_th}, the spectrum produced by applying $W^{\rm l2}_{n} (l^{\prime})$ indeed agrees well with the experiment. Hence, by including fine-tuning on $l-$distribution, the current CX model can explain the EBIT result. We emphasize that the $n$-scaling used for the modeling has an uncertainty which is larger than our experimental determination of the centroid of the series limit. For an atomic hydrogen CX donor, the centroid of the observed spectra could be shifted by perhaps one unit of the principal quantum number $n$.

\newpage

%
%
\subsection{Comparison with the astrophysical observations}
%
%

Following our experimental finding of a set of transitions from highly excited states in bare sulfur ions in agreement with model predictions of Gu and earlier work on these \cite{martinez2014} and other HCI, we compare our modeled data considering atomic H as a CX donor with some of the recent astrophysical observations. 

\begin{figure}
	\includegraphics[width=0.95\columnwidth]{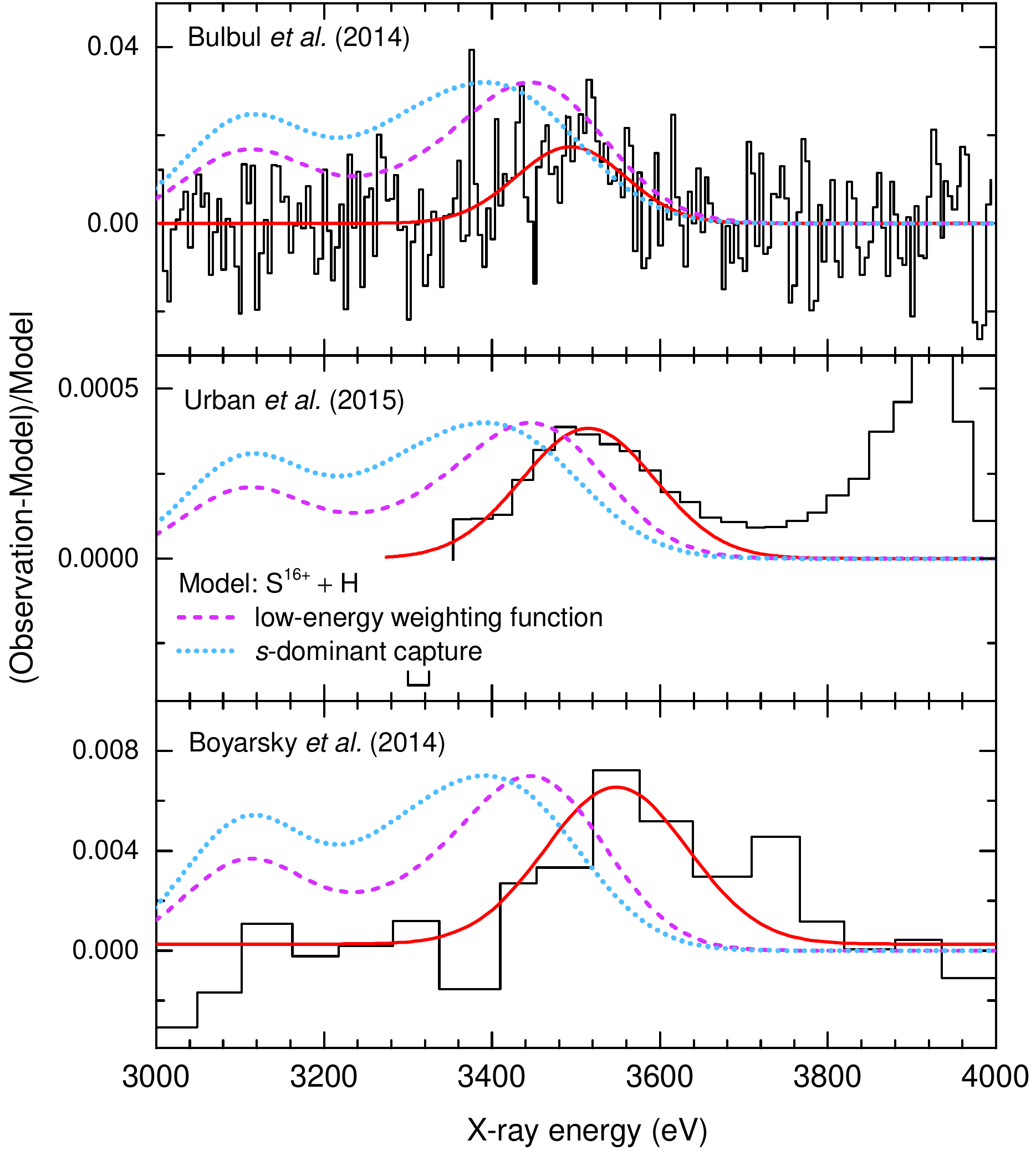}
	\caption{Astrophysical observations reported in top right panel of Fig.~5 in~\citet{bulbul2014}, left panel of Fig.~2 in~\citet{urban2015}, and right panel of Fig.~1 in~\citet{boyarsky2014} are shown. The red solid curves represent the Gaussian line with the energies fixed to the reported values. Synthetic spectra ($\mathrm{S}^{16+} + \mathrm{H}$) using low-energy weighting (dashed line) and with \textit{s}-dominant capture (dotted line) functions are compared the observations.\label{fig:astro_obs}
}
\end{figure}

Due to the weak signal and other sources of uncertainty in the modeling of the astrophysical X-ray spectrum, there are uncertainties for the photon energy of ULF. Its centroid energy has been variously reported: 3.46 -- 3.53 keV in the observation of the Andromeda nebula and the Perseus galaxy cluster ~\citep{boyarsky2014}, 3.51 -- 3.57 keV in the stacked spectra of 73 galaxy clusters~\citep{bulbul2014}, 3.54 keV in the Galactic center~\citep{boyarsky2015} and 3.51 -- 3.59 keV in the Perseus cluster~\citep{urban2015}. We took the spectral fitting residuals reported from the observations for comparison with our CX model, see Fig.~\ref{fig:astro_obs}. These spectral residuals are produced by fitting the observed data with thermal plasma models. 
%
%
We note that the expected $n=3\rightarrow 1$ transition in $\mathrm{S}^{15+}$ at 3.1 keV does not show up in the data sets taken from the papers by Bulbul, Urban and Boyarski~\citep{bulbul2014,urban2015,boyarsky2014} that are displayed in Fig.~\ref{fig:astro_obs}. This would contradict with our CX model, which predicts that line. An explanation for this was already given in~\citet{gu2015}: Basically, it rests on the fact that it would be extremely difficult to detect any excess of CX at any photon energy for which strong transitions are expected. At the position of the $\mathrm{S}^{15+}$ $n=3\rightarrow 1$ line there is a blend with the strong $K_\alpha$ transition of heliumlike $\mathrm{Ar}^{16+}$. The spectral models used by Bulbul, Urban and Boyarski are adjusted to yield a zero photon excess at 3.12 keV (Ar \textsc{xvii} $n=2\rightarrow 1$ 'triplet' lines). Moreover, the models used include~\citep{bulbul2014,urban2015,boyarsky2014} further strong Ar, S, and Ca lines in the 3--4 keV range that were independently fitted. These free fits explain why the sulfur transition n=3 to 1 is seemingly absent from the x-ray photon excess data. Furthermore, if those models slightly overestimate the contribution from hydrogenlike $\mathrm{Ar}^{17+}$ Lyman-$\alpha$ (Ar \textsc{xviii} at 3.31 keV), one would expect that the subtracted photon excess signal would experience a shift of its centroid, resulting in an ULF at an apparent energy higher than its actual value. Therefore, we conclude that even comparatively small uncertainties of the spectral models used by Bulbul, Urban and Boyarski can certainly explain the minor shift to higher energies of the ULF in comparison with our CX modeled spectrum.

%
%
\section{Conclusion}
%
%

We have presented experimental charge exchange spectra of $\mathrm{S}^{16+}$ and $\mathrm{S}^{15+}$ interacting with a $\mathrm{CS}_2$ gas target in an electron beam ion trap, and found excellent agreement with analogous work by the LLNL group~\citep{martinez2014}. The inferred energy of a charge-exchange-induced spectral feature 3.47 $\pm$ 0.06 keV is in full accord with both the astrophysical observations and with our own FAC calculations, and confirms the prediction of Gu and Kaastra. The transitions observed appear at photon energies which are statistically consistent with the astrophysical observations at the level of mutual uncertainties. We conclude that inclusion of CX in the modeling of the astrophysically observed spectra with the inclusion of uncertainties given by the lack of experimental data on charge exchange of HCI with atomic hydrogen might well explain a large fraction, and perhaps even all of the photon signal observed from galaxies and clusters at $\sim$3.5 keV. Furthermore, we cannot see a reason why this well-known process should not be active in the cases that have been purported, since both the cosmic abundances of sulfur and hydrogen compellingly point to its ubiquity in the neighborhood of galactic winds. Indeed, other explanations are possible, including some K~\textsc{xvii} lines and low-energy satellites of~\emph{KLM} dielectronic recombination line in Ar~\textsc{xvii}. However, charge exchange is a proven mechanism and its exclusion would require some additional explanation which has not been yet put forward. The absence of CX-induced X-ray transitions from high-$n$ Rydberg states in models and databases should, therefore, be corrected, since the inclusion of this process certainly would  improve the accuracy of the spectral models.  
%

%
%

\bibliographystyle{yahapj}
\bibliography{bibliography}

%
%

\end{document}